\documentclass[twocolumn,showpacs,preprintnumbers,amsmath,amssymb,prl,epsf,superscriptaddress]{revtex4}
\usepackage{graphicx}
\usepackage{nicefrac}
\usepackage[squaren]{SIunits}

\begin{document}
\title{Two-Color Bright Squeezed Vacuum}
\author{Ivan N. Agafonov}
\affiliation{Department of Physics, M.V.Lomonosov Moscow State
University, \\ Leninskie Gory, 119992 Moscow, Russia}
\author{Maria~V.~Chekhova}
\affiliation{Department of Physics, M.V.Lomonosov Moscow State
University, \\ Leninskie Gory, 119992 Moscow, Russia}
\affiliation{Max Planck Institute for the Science of Light,
G\"unther-Scharowsky-Stra\ss{}e 1/Bau 24, 91058 Erlangen, Germany}
\author{Gerd Leuchs}
\affiliation{Max Planck Institute for the Science of Light, G\"unther-Scharowsky-Stra\ss{}e 1/Bau 24,
91058 Erlangen, Germany}
\affiliation{University Erlangen-N\"urnberg, Staudtstrasse 7/B2, 91058 Erlangen, Germany}
\begin{abstract}
In a strongly pumped non-degenerate traveling-wave OPA, we produce
two-color squeezed vacuum with up to millions of photons per pulse.
Our approach to registering this macroscopic quantum state is direct
detection of a large number of transverse and longitudinal modes,
which is achieved by making the detection time and area much larger
than the coherence time and area, respectively.  Using this
approach, we obtain a record value of twin-beam squeezing for direct
detection of bright squeezed vacuum. This makes direct detection of
macroscopic squeezed vacuum a practical tool for quantum information
applications.
\end{abstract}
\pacs{42.50.Lc, 42.65.Yj, 42.50.Dv}
 \maketitle \narrowtext
\vspace{-10mm}

We are now witnessing a growth of interest in macroscopic quantum
systems~\cite{macroqubits}. Although the subject of macroscopic
superpositions is causing scientific debates ever since the
formulation of the famous Schroedinger-cat paradox, nowadays the
interest in macroscopic quantum systems is also motivated by
applications: gravitational wave detection~\cite{Lam}, quantum
memory~\cite{Polzik}, super-resolution~\cite{Dowling}, etc. An
interesting perspective for applying macroscopic nonclassical states
of light in quantum technologies stems from the fact that such
states, in principle, can provide stronger interactions with matter
than microscopic (single-photon and few-photon) states.

An evident example of a macroscopic quantum state of light is
two-mode bright squeezed vacuum (SV) generated at the output of a
traveling-wave optical parametric amplifier (OPA)~\cite{MandelWolf}.
At weak pumping such an OPA produces biphoton light, which is
usually characterized in terms of  Glauber's normalized correlation
functions and studied using single-photon coincidence counting
technique. At strong pumping, two-photon correlations become smeared
by the presence of higher photon numbers and coincidence counting
gets inefficient since normalized Glauber's correlation functions
approach unity. This does not mean that quantum correlations
disappear; they just need a different type of measurement to be
revealed~\cite{JETPLett}. This other kind of measurement can be
based on the photon-number difference. Since signal and idler beams
are 'twins'~\cite{unseededOPA}, their photon-number difference does
not fluctuate, and its variance is ideally zero. When such beams are
detected, the difference of the  output photocurrents of the two
detectors fluctuates only due to non-unity quantum efficiencies of
the detectors, electronic noise, or possible optical losses.

Twin-beam squeezing for two-mode SV is being studied using optical
homodyne technique for two decades (see, e.g.,
Refs~\cite{Levenson,Vasilyev,Schnabel}). The method is powerful
since it provides phase information. At the same time, mixing with
local oscillator turns SV into a different state, namely a squeezed
coherent state. Obtaining the 'unperturbed' properties of SV
requires direct detection. The latter has been applied to bright SV
rather recently~\cite{Lugiato,Bondani,Brida2}, with the appearance
of sensitive CCD cameras and the technique of charge-integrating
detection.  All these pioneering works showed rather low squeezing
(less than 3dB), which disappeared at large photon numbers (more
than 10 photons per mode). As we will show below, most probable
reason for this behavior is not selecting enough transverse modes.
There was, however, one very early paper on the direct detection of
weak two-mode SV ~\cite{Tapster} cleverly using multimode collection
and obtaining a certain degree of squeezing.

Measurements based on the photocurrent subtraction are principally
different from the ones based on coincidence counting (or
photocurrent multiplication). In particular, while coincidence
measurement requires as few modes selected as possible, measurement
of the difference-signal variance requires a large number of
detected modes. (Alternatively, a single Schmidt mode can be
selected in both signal and idler beams~\cite{Schmidt,Rytikov}.)
Recently, $3$dB of polarization squeezing was observed via direct
detection of SV by collecting a large number of frequency and
angular modes~\cite{erlsqueezing}, although with the values of gain
as small as $0.3$. In this work, we measure two-mode squeezing for
two-color SV generated via collinear nondegenerate type-I parametric
down-conversion at gain values of up to four.

The effect of two-mode squeezing consists of the suppression of
fluctuations in the photon-number difference for two light modes
(beams) below the classical limit, which is given by the mean sum
photon number in these beams. Indeed, it is easy to show that for
two single-mode beams, labeled $1,2$, with mean photon numbers equal
to $N$,  the variance of photon-number difference is
\begin{equation}
\hbox{Var}(\hat{N}_1-\hat{N}_2)=N^2(g_{11}^{(2)}+g_{22}^{(2)}-2g_{12}^{(2)})+2N,
\label{Var}
\end{equation}
where $g_{11}^{(2)},g_{22}^{(2)}$ are normalized second-order
intensity correlation functions for beams 1 and 2, respectively, and
$g_{12}^{(2)}$ is their second-order cross-correlation function. For
classical light beams, according to Cauchy-Schwarz inequality,
$g_{11}^{(2)}+g_{22}^{(2)}-2g_{12}^{(2)}\ge 0$~\cite{Walls}, and
$\hbox{Var}(\hat{N}_1-\hat{N}_2)\ge 2N$. Violation of this
inequality can be considered as a sign of nonclassicality. Hence, a
usual measure for twin-beam squeezing is Noise Reduction Factor
(NRF)~\cite{Aytur,Lugiato}:
\begin{equation}
\hbox{NRF}=\hbox{Var}(N_1-N_2)/2N. \label{NRF}
\end{equation}
For two-mode SV, $g^{(2)}_{11}=g^{(2)}_{22}=2$ (both twin beams have
thermal statistics), $g^{(2)}_{12}=2+1/N$, and, according to
Eq.~(\ref{Var}), NRF=0.

Our setup was a traveling-wave OPA based on two type-I 2-mm BBO
crystals with the optic axes oriented in the same plane but
symmetrically with respect to the pump direction, in order to
eliminate spatial walkoff. As a pump, we used the third harmonic of
a Nd:YAG laser (wavelength $\lambda_p=355$nm) with pulse duration 17
ps, repetition rate 1 kHz, and energy per pulse up to 0.2 mJ. The
fundamental and second-harmonic radiation of the laser was
eliminated using a prism. The crystals were oriented to produce
frequency-nondegenerate parametric down-conversion (PDC) at
wavelengths $635$ nm (signal) and $805$ nm (idler). After the
crystals, the pump radiation was cut off by two dichroic mirrors
with high transmission ($T>97\%$) at the down-converted wavelengths.
The residual pump radiation was suppressed by two Glan prisms, the
first one selecting the pump polarization (H) before the crystals
and the second one transmitting the SV polarization (V). The pump
power could be varied by rotating a waveplate before the first Glan
prism. Signal and idler beams were then separated using a dichroic
beamsplitter and focussed, by lenses with 5 cm focal lengths, on two
charge-integrating detectors based on Hamamatsu S3883 p-i-n
diodes~\cite{Hansen,erlsqueezing}. Quantum efficiencies of the
detectors at wavelengths $635$ nm and $805$ nm were $85$\% and
$95$\%, respectively. The angular bandwidth selected was determined
by the sizes of two iris apertures (about $1$ cm) placed at a
distance of $36$ cm from the crystals. (In order to properly match
frequency and angular modes in the signal and idler channels, the
apertures had different diameters, see below.) The number of
transverse modes $m_t$ was given by the squared ratio of the signal
(idler) aperture size and the transverse coherence length of signal
(idler) radiation in the plane of the aperture (about $2$ mm),
yielding $m_t\approx25$. The number of longitudinal modes was given
by the ratio of the pulse duration and PDC coherence time
(corresponding to the spectral bandwidth selected by the apertures,
about $13$ nm), yielding $m_l\approx150$. The overall detection
efficiencies at wavelengths $805$ nm and $635$ nm, determined by 11
AR coated surfaces, 3 dichroic coatings and detectors' quantum
efficiencies, were estimated to be $77$\% and $70$\%, respectively,
resulting in the best achievable NRF of $0.33$.
\begin{figure}
\includegraphics[width=0.38
\textwidth]{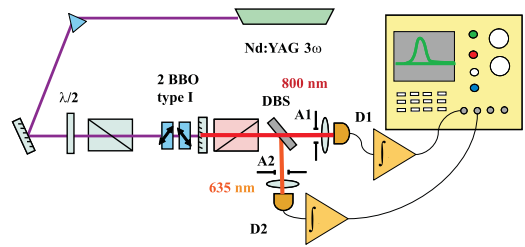} \caption{(Color online) Experimental setup.
DBS, dichroic beamsplitter; D1 and D2, detectors; A1 and A2,
apertures.}
\end{figure}
Registration of the detectors' output signals and their calibration
was performed the same way as described in Ref.~\cite{erlsqueezing}.
However, in comparison with the results of~\cite{erlsqueezing} we
now had much larger photon numbers, on the order of $10^6$ photons
per pulse, and
 the shot-noise level (about $10^3$ electrons) much
exceeded the electronic noise level ($180$ electrons). Still, in the
results shown below the electronic noise was subtracted.

The most important point about direct detection of SV is that the
detection time and detection area should be much larger than the
coherence time and coherence area, respectively. In other words, the
number of detected modes should be large, as it was shown
in~\cite{Brambilla,Rytikov}. In particular, the detection aperture
size should much exceed the transverse coherence length of
down-converted radiation.
%In experiments with CCD
%cameras~\cite{Lugiato,Brida2,Brida}, this condition could not be
%satisfied because of the small size of a single pixel.
In order to
study the dependence of squeezing on the number of registered modes
we made the measurement for different sizes of the apertures. First,
signal and idler aperture diameters $D_s,D_i$ were varied
simultaneously, so that their ratio satisfied the relation (set by
the transverse phase matching condition for PDC)
\begin{equation}
D_i/D_s=\lambda_i^{max}/\lambda_s^{min}, \label{apertures}
\end{equation}
where $\lambda_{i}^{max},\lambda_{s}^{min}$  are maximal idler and
minimal signal wavelengths selected, respectively (Fig.2a). Then,
the idler aperture had a fixed diameter and the signal aperture
diameter was varied (Fig.2b).
\begin{figure}
\includegraphics[width=0.28\textwidth]{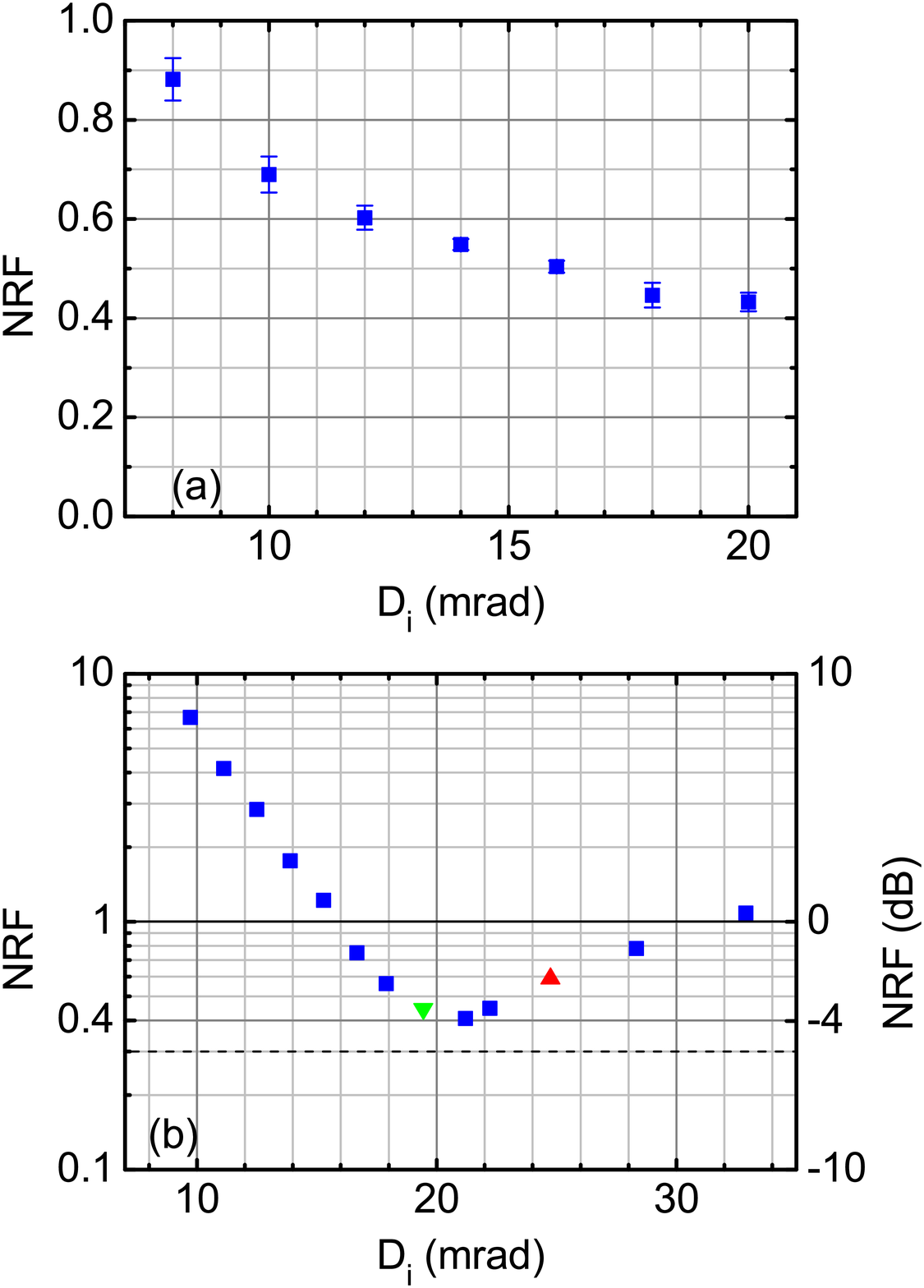}
\caption{(Color online) Dependence of NRF on the aperture angular
sizes for (a) simultaneous variation of both aperture diameters and
(b) variation of only signal aperture, the idler one remaining fixed
at the position marked by red 'up' triangle. The green 'down'
triangle shows the size given by condition~(\ref{apertures}). Dashed
line corresponds to the NRF value calculated from the specifications
of the optical elements.}
\end{figure}
From Fig.2a we see that, according to the predictions
of~\cite{Brambilla}, squeezing improves with the increase of the
aperture diameters, i.e., with the number of angular modes selected.
This can be understood by recalling that squeezing is sensitive to
losses, so that collection of a limited number of modes leads to the
existence of uncorrelated photons of the beam, which is similar to
the loss of a certain amount of photons. Ideally, to observe maximal
squeezing one should detect all frequency-angular spectrum of PDC.
But, as it was shown in Ref.~\cite{Rytikov}, in practice a high
degree of squeezing can be obtained even for a limited number of
modes provided that it is high.

Fig.2b shows that the optimal squeezing is obtained when aperture
sizes are matched according to condition (\ref{apertures}). If the
aperture sizes are not matched, squeezing disappears and even turns
into anti-squeezing. This happens because both signal and idler
beams are thermal ones, and they consist of strongly fluctuating
'speckles'. There is of course perfect correlation between the
speckles in the two beams. If the aperture sizes are not matched,
then in one of the beams there are 'unmatched' speckles whose excess
intensity fluctuations are not compensated and lead to an increase
in the NRF. A straightforward calculation shows that for two sets of
$k$ independent thermal modes with equal photon numbers
$\mathcal{N}$, noise reduction factor does not depend on $k$ and is
only determined by the mean photon number per mode:
$\hbox{NRF}=\mathcal{N}+1$. In other words, difference-photocurrent
measurement on a large number of independent thermal modes will
provide NRF as high as for a single thermal mode. This is yet
another demonstration of the huge difference between NRF measurement
and photocount coincidence detection where thermal beams
'poissonise' if a large number of modes is collected. The
discrepancy between the experimental NRF value of 4 dB and the
calculated value of 5.2 dB might be due to imperfect alignment
(apertures shape deviating from round etc).

It follows that for bright PDC, mismatch of the aperture diameters
will be more noticeable at large numbers of photons per mode, i.e.,
at large gain. In fact, whether two-mode squeezing can be still
observed for SV at high gain is an important question. Previous
works~\cite{Lugiato,Brida} showed an increase of NRF at high values
of the gain. At the same time, in theory, NRF for SV should  only be
given by losses and not depend on the gain. The growth of NRF with
the increase of parametric gain was explained in
Refs.~\cite{Lugiato,Brida} by the dependence of the mode size on the
gain, and, as a consequence, the reduction of the selected number of
modes with the gain growth. However, this should have little effect
if a large number of modes are collected; also, this does not
explain the appearance of anti-squeezing.

To study the dependence of squeezing on the gain (Fig.3a), we
measured NRF versus the pump power. The aperture sizes were matched
in accordance with Eq.(\ref{apertures}). The gain was estimated by
fitting the dependence of the PDC output photon number $N$ on the
pump power (shown in the inset to Fig.3a) with the formula
\begin{equation}
N=m\sinh^2\Gamma+N_0, \label{sinh}
\end{equation}
where $\Gamma$ is the parametric gain, scaling as square root of the
pump power, $m=m_t m_l$, and $N_0$ is the noise (fluorescence)
linear in the pump, measured separately.
\begin{figure}
\includegraphics[width=0.28\textwidth]{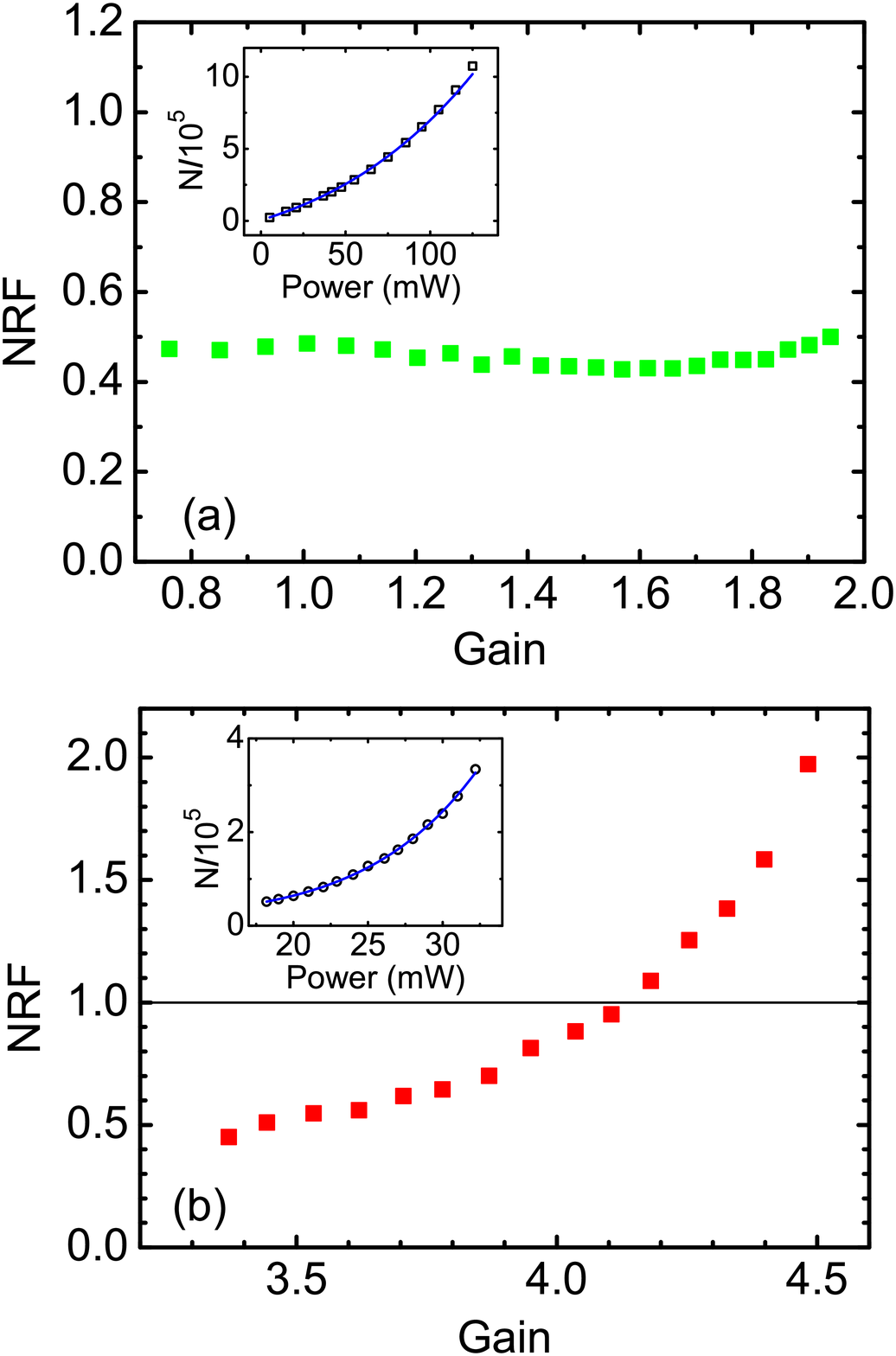}
\caption{(Color online) NRF versus the gain for (a) unfocused and
(b) focused pump. Insets: output photon numbers versus the pump
power.}
\end{figure}
In all dependencies, NRF indeed grew  with the increase in the gain.
However, this growth was very sensitive to alignment. For the best
alignment we could achieve, a growth of NRF is only noticeable at
the end of the dependence, where $\Gamma\approx 2$ (Fig.3a).
%To analyze the origin of such
%behavior, we checked if it could be possibly caused by the
%fluctuations in the pump intensity (3\%, according to the laser
%specifications). These fluctuations were eliminated by means of
%postselection,  by independently measuring amplitudes of the pump
%pulses and choosing only those within the shot-noise limit. The data
%obtained for postselected pulses indeed shows a certain decrease in
%the NRF but this decrease is very small (green solid triangles in
%Fig.3).

To pass to higher-gain PDC, we focused the pump into the crystals by
means of a 5:1 telescope. This way we achieved gain values of up to
$4.5$, where the increase of NRF was noticeable (Fig.3b). Still,
even at $\Gamma=4$, corresponding to about $700$ photons per mode,
squeezing was still observed. At high gain, the sensitivity of
squeezing to misalignment was even more evident. For instance,
Fig.4a shows NRF as a function of the mean photon number per mode
$\mathcal{N}= \sinh^2\Gamma$ for the best alignment we could achieve
(squares) and for the idler aperture (whose diameter was $10.2$ mm)
displaced by $500$ $\mu$m (circles). Clearly, this 5\% displacement
of the aperture leads to a dramatic increase in NRF. Fig.4b shows
NRF as a function of the aperture shift measured at the gain value
$3.7$.
\begin{figure}
\includegraphics[width=0.28\textwidth]{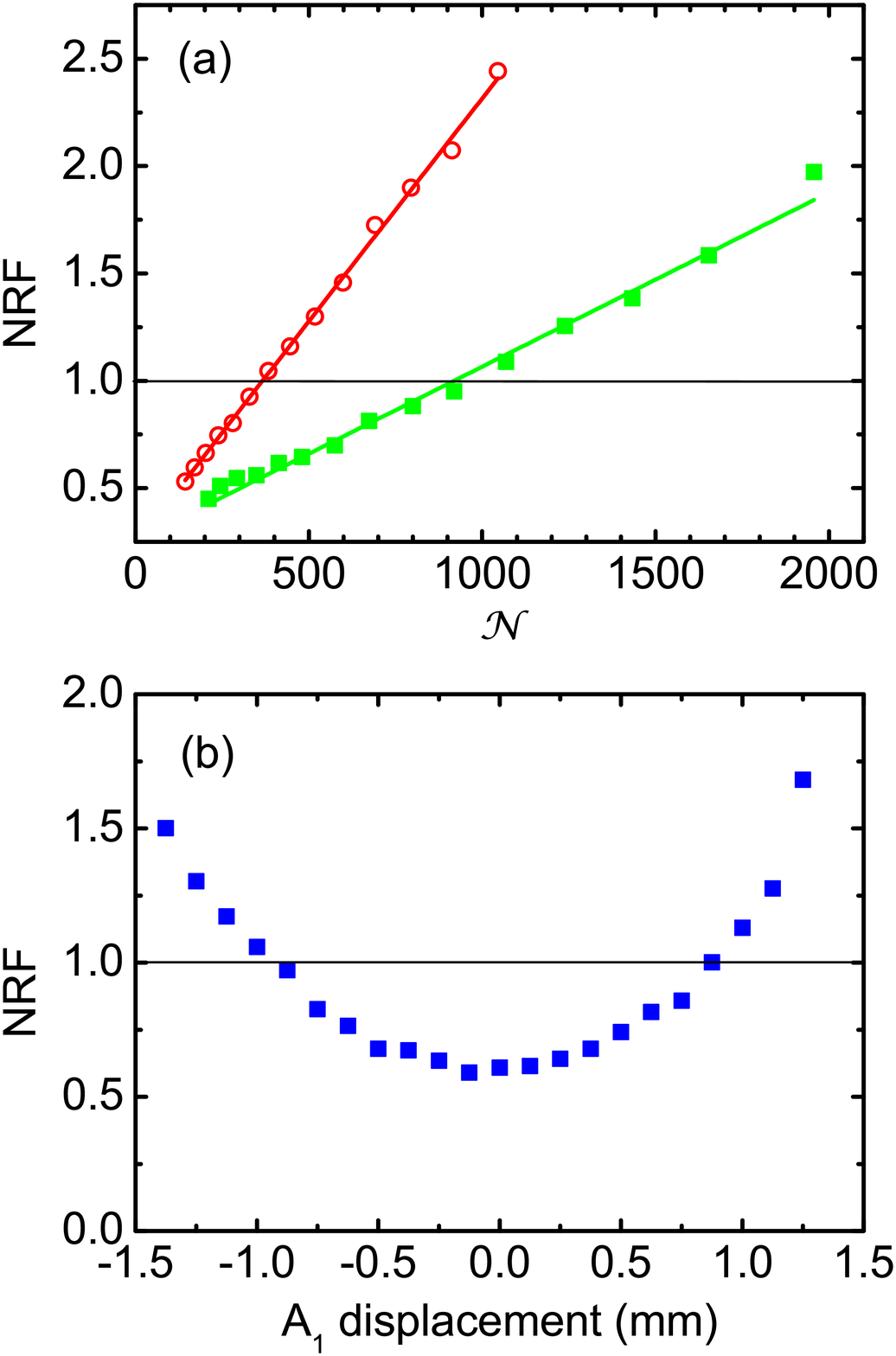}
\caption{(Color online) The effect of aperture shift on the NRF. (a)
Solid green squares: best alignment, red empty circles: aperture in
the idler channel shifted by $500$ $\mu$m. (b) NRF versus the
aperture displacement, measured at $\Gamma=3.7$.}
\end{figure}

High sensitivity of NRF to misalignment is of the same origin as its
sensitivity to the aperture size mismatch. Indeed, if the mode sets
collected by the signal and idler detectors are not matched, there
are uncompensated (independent) intensity fluctuations in the signal
and idler channels. They should lead to NRF growth, scaling linearly
with the mean photon number. Our calculation shows that, provided
that in each channel there are $m$ matched PDC modes and $k$
unmatched ones, all having the same mean photon number
$\mathcal{N}$, NRF is increased by
\begin{equation}
\Delta=\frac{k(\mathcal{N}+1)}{m+k}. \label{unmatched}
\end{equation}
For small displacements of a circular aperture, this should lead to
a linear dependence of NRF on $\mathcal{N}$, which we used to fit
the experimental data in Fig.4a.

It follows that observation of twin-beam squeezing in high-gain PDC
is the more sensitive to mode matching the higher the gain. This
explains a linear growth of NRF with the increase of $\mathcal{N}$
observed in~\cite{Lugiato}. Similar effect is used in quantum
metrology to measure spatial displacements with high
precision~\cite{Treps}, and determines the resolution of quantum
imaging~\cite{imaging}.

%Finally, we tested the sensitivity of the obtained squeezing to
%optical losses. Optical losses were introduced by placing a halfwave
%plate before the second Glan prism and rotating it so that the
%transmission of the optical system for the down-converted light
%varied by about $10$ dB. Because rotation of the half-wave plate
%also changed the efficiency of the pump suppression, we had to
%additionally insert a long-pass filter after the Glan prism. The
%measurement showed a linear dependence of NRF on the optical
%transmission, in full agreement with the theoretical prediction.
%\begin{figure}
%\includegraphics[width=0.3\textwidth]{Fig5.eps}
%\caption{NRF as a function of optical transmission. Line:
%theoretical fit $\hbox{NRF}=1-0.52T$.}
%\end{figure}

In conclusion, we have generated and tested via direct detection
two-color squeezed vacuum containing up to a thousand photons per
mode, and up to a million of photons per pulse. By collecting a
large number of angular and frequency modes in both channels,
considerable amount of two-mode squeezing (4dB) has been obtained,
the best result ever achieved for the direct detection of squeezed
vacuum. We have also studied the behavior of squeezing on the
angular bandwidths selected in signal and idler channels and showed
that the effect is extremely sensitive to the sizes and alignment of
the angle-selecting apertures. On the other hand, if a large number
of conjugated angular modes is properly selected, squeezing is not
much sensitive to the growth of the parametric gain. In particular,
we observed nearly constant amount of squeezing up to the gain
$\Gamma =2$, corresponding to $13$ photons per mode. Squeezing,
although to a smaller degree, was observed up to $900$ photons per
mode.

Our result is an important step in the investigation of mesoscopic
and macroscopic quantum systems. We show that
difference-photocurrent measurements can replace coincidence-based
measurements in the study of macroscopic squeezed vacuum. Pairwise
correlations are still evident provided that losses are sufficiently
small and proper mode matching is ensured.

We acknowledge the financial support of the European Union under
project COMPAS No. 212008 (FP7-ICT) and the Russian Foundation for
Basic Research, grants \#08-02-00741a and \#10-02-00202a. I.N.A.
acknowledges the support of the 'Dynasty' foundation and M.V.C., the
support of the DFG foundation.


\begin{references}
\bibitem{width}When calculating the aperture sizes we did not take
into account finite spectral bandwidth of PDC at fixed angle of
emission.
\end{references}
\end{document}